\documentstyle[psfig]{mn}
\newcommand{\be}{\begin{equation}}
\newcommand{\ba}{\begin{eqnarray}}
\newcommand{\ee}{\end{equation}}
\newcommand{\ea}{\end{eqnarray}}

\def\hn{{\bf \hat n}}
\def\hnp{{\bf \hat n}'}

\begin{document}
\title{Temperature Correlations in a Finite Universe}
\author[Evan Scannapieco, Janna Levin, and Joseph Silk]
{Evan Scannapieco,${}$\thanks{\tt evan@astro.berkeley.edu}
 Janna Levin,${}$\thanks{\tt janna@cfpa.berkeley.edu}
 and Joseph Silk${}$\thanks{\tt silk@pac2.berkeley.edu}\\
Center for Particle Astrophysics, UC Berkeley, Berkeley, CA 94720-7304.}

\maketitle

\begin{abstract}
We study the effect of a finite topology on the
temperature correlations of the cosmic microwave background in a flat 
universe.  Analytic expressions for the angular power spectrum are 
given for all possible finite flat models.
We  examine the angular correlation function itself,
pointing out visible and discrete
 features that arise from topology.
While observations of the power spectrum on large angular scales can 
be used to place bounds on the minimum
topology length, cosmic variance generally restricts us from
differentiating one flat topology from another. 
Schemes that acknowledge anisotropic structures, 
such as searches for ghosts, circles or geometric patterns,
will be needed to further probe topology.
\end{abstract}


\section{Introduction}

There is no reason to believe that the universe is infinite.
While general relativity specifies the local curvature of spacetime,
the global geometry of the universe remains unspecified.  From this
point of view, an infinite universe is assumed
only to simplify theoretical calculations and is subject to 
observational verification.
Although inflation would push topology scales
far out of view, recent observations \cite{open2,open1} suggest 
that the 
curvature of the universe may deviate from flat sufficiently to be 
measurable.
If curvature is observable, then how can we assume topology is not?

As it represents the largest volume observable, the Cosmic Microwave
Background (CMB) is uniquely sensitive to the global geometry of the 
universe.  Already the Cosmic Background Explorer 
(COBE) results have been used 
to place constraints on flat topologies and limited open topologies
that are orders of magnitude better than other 
approaches \cite{flat1,flat2,sss,star,lum1,lum2,costa,lbbs,us}, 
hereafter LSS).
The increased sensitivity of 
the next generation of
CMB experiments has
inspired renewed interest in the search for topology
\cite{css1,css2,bps,lssb}.

In LSS, we were able to quantify
the effects of topology
on the CMB by considering all possible compactifications of flat
space.  We solved for the spectrum of 
fluctuations explicitly which allowed us to create
realizations of
finite universes and compare typical angular power spectra to the 
COBE data.
Here we extend those results by computing
the ensemble-averaged angular power spectrum, as opposed to just
obtaining realizations.  
Since we know the modes explicitly from LSS, our task here is 
to reduce the angular power spectrum to a simple 
analytic expression for each of the six orientable topologies.
Generic features in the spectrum can then be identified 
without ambiguity.

\section{Temperature Fluctuations}

The primary cause of CMB temperature fluctuations is
lumps in the geometry of spacetime at the time of decoupling.
The fluctuations can be decomposed into eigenmodes and written
in any compact, flat
spacetime as
        \be
        {\delta T\over T}(\hat n)\propto \sum_{\vec k}
        \hat \Phi_{\vec k}
        \exp \left(i\Delta \eta \vec k \cdot \hn \right)
         ,\label{eq1}
	\ee
with $\Delta \eta $ the conformal time between today and 
decoupling. 
On a compact manifold, the usual
continuous spectrum of eigenvalues
becomes discretized, hence the sum in \ref{eq1}.
The $\hat \Phi_{\vec k}$ are primordially seeded
Gaussian amplitudes
that obey the reality condition
$\hat \Phi_{\vec k}=\hat \Phi_{-\vec k}^*$ and a set of relations
that depend on the topology (LSS).

With this decomposition we can construct the
correlation function between any two points on the sky as
\ba
&C&(\hn,\hnp)\,  = \,
\left\langle {\delta T\over T}(\hn) {\delta T\over T}(\hnp) \right\rangle \\
\nonumber
&\propto&
\sum_{\vec k}\sum_{\vec k'}
\left\langle \hat \Phi_{\vec k} \hat \Phi_{\vec k'}^* \right\rangle
       \exp\left(i\Delta \eta (\vec k \cdot \hn - \vec k' \cdot \hnp ) 
\right ).
\label{eq:cnn}
\ea
As the fundamental domain has a particular orientation on the sky, 
the correlation is not simply a function of the angular separation
between $\hn$ and $\hnp$ as it is in the infinite case.

From this expression,
$C_\ell$ can be determined using the orthogonality relations of the
Legendre polynomials:
\be
C_\ell 
%= 2 \pi \int_{-1}^1 d \mu C(\arccos \mu) P_\ell(\mu)
= \frac{1}{4 \pi} \int d \Omega \int d \Omega' C(\hn, \hnp) P_\ell(\mu),
\ee
where $\mu = \hn \cdot \hnp$.
Expanding the exponential and 
Legendre polynomials in terms of spherical harmonics, this becomes
\ba
C_\ell \propto
\sum_{\vec k} \sum_{\vec k'}
&\left\langle \hat \Phi_{\vec k} \hat \Phi_{\vec k'}^* \right\rangle
 {j_\ell(\Delta \eta k) j_\ell(\Delta \eta k')}(2 \ell + 1)^{-1} \\ \nonumber
& \times \sum_{m = -\ell}^{\ell}
Y_{\ell,m}^*({\bf \hat k}) Y_{\ell,m}({\bf \hat k'}) .
\label{eq:cl}
\ea
As $\left\langle \hat \Phi_{\vec k} \hat \Phi_{\vec k'}^* \right\rangle$
and the spectrum of eigenvalues are known for all six possible flat
topologies, we can use this expression to compute $C_\ell$ for
each of the possible cases.

\section{Compact Spaces}

\begin{figure}
\centerline{\psfig{file=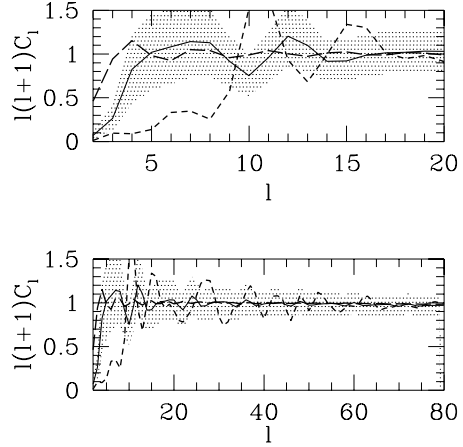,width=2.4in}}
\caption{Normalized $C_\ell$s for the torus with the
topology scale equal to $0.66$ (short dashed) $1.0$ (solid) and 
$1.5$ (long dashed) times the radius 
of the horizon.  Estimated cosmic variance about the unit torus is
shaded.  The upper panel is a detail of low $\ell$s.
The lower panel shows a wider range of $\ell$s.}
\label{fig:torus}
\end{figure}

The simplest topology is the hypertorus, which is  
built out of a parallelepiped
by identifying $(x,y,z)\rightarrow (x+h,y+b,z+c)$.
The identification leads to a restriction of the eigenvalue spectrum,
$\vec k=2\pi(j / h, w/b,n/c)$
with the $j,w,n$ running over all integers.
With this restriction, the $C_\ell$s become
\be
C_{\ell} 
%\propto
%\sum_{jwn} \frac{{\cal P}(k)}{k^3}
%\frac{j_\ell(\Delta \eta k)^2}
%{2 \ell + 1} \sum_{m = -\ell}^{\ell}
%|Y_{\ell,m}({\bf \hat k})|^2
\propto \sum_{jwn} \frac{{\cal P}(k)}{k^3} j_\ell(\Delta \eta k)^2,
\label{eq:dan}
\ee
where ${\cal P}(k) \propto 1$ for a flat power spectrum. 
This is in agreement with \cite{sss}.
In Fig.\ \ref{fig:torus} we plot this expression for three different topology
scales for a flat power spectrum normalized by what we would expect
for a universe with no topology; that is $C_\ell \times
{\rm V}^{-1} (2 \pi)^2 \ell (\ell+1)$, where $V$ is the volume of the 
fundamental domain.  The normalization is absolute, such
that an infinite universe would be represented by    
normalization $\ell (\ell+1) C_\ell = 1.$  Cosmic variance is estimated
as $C_\ell \sqrt{2/({2 \ell +1})}$, as for an infinite 
universe, although the true variances for any given topology would be 
slightly different.

There are a number things to note here.  
In the upper panel we see that the low $\ell$ modes are 
damped, with the suppression becoming more severe as the topology scale 
decreases.
From the discretization of the 
wave vector, $\vec k$, it is clear that there is a minimum eigenvalue
corresponding to the longest wavelength that can fit inside the fundamental
domain.  This maximum wavelength can be associated with an
angular scale above which we do not expect to find fluctuations. 
As the association between real space and angular perturbations causes some
averaging over $k$ modes, the damping is smeared over a range of 
$\ell$ values.

In addition to the damping at low $\ell$, a finite topology
also causes jags at higher $\ell$ values,
extending to values above $\ell = 60$ for the torus 
of size $.66 \Delta \eta$.  The
jaggy features not only suppress 
many of the higher
$C_\ell$s but actually cause {\em enhancement} at selected $\ell$ values.
This ringing in the $C_\ell$s can be understood as caused by the presence
of a discrete set of harmonics of the fundamental domain in the matter power
spectrum.  The discretization not only draws power away from values that
are disallowed, but enhances power at certain typical angular scales.
Another way to understand this effect is to consider the presence
of multiple copies of the same point.
One can imagine that given a topology scale, there are certain angles
at which multiple images tend to fall, while at other angles
such correlations are disallowed by the geometry of the fundamental domain.
This effect will become more apparent when we consider the angular 
correlation function itself.

\begin{figure}
\centerline{\psfig{file=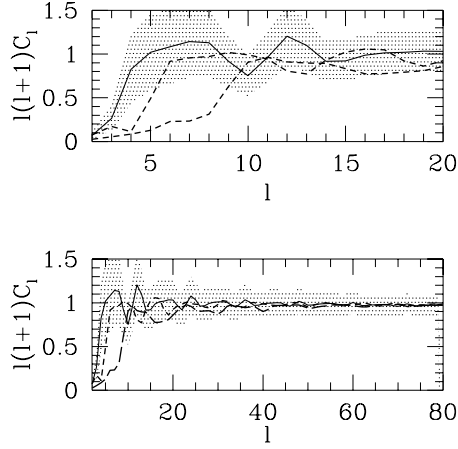,width=2.4in}}
\caption{Normalized $C_\ell$ for the torus with unit volume
and three different aspect ratios: h=b=c (solid), h=2b=2c (short dashed)
h = 2b = 3c (long dashed), 
again with estimated cosmic variance about the torus.
The upper panel is a detail of the lower panel.}
\label{fig:aspect}
\end{figure}

In Fig.\ \ref{fig:aspect} we plot the $C_\ell$s for three different tori
with the same volume but different aspect ratios.
Here we see that the suppression at low $\ell$s is much more severe 
in the elongated configurations than in the model with equal sides.
Thus the damping is much more dependent on the minimum 
dimension of the parallelepiped than the overall volume of the fundamental
domain.

Returning to Eq.\ \ref{eq:cl},
we are able to calculate the effect of a finite volume on the $C_\ell$s for
more complicated flat topologies.
Three other spacetimes are constructed from a
parallelepiped.
The first twisted parallelepiped we consider 
has opposite faces identified with one pair rotated through the angle $\pi $.
The eigenmodes are 
$\vec k=2\pi(j / h, w/b,n/2c)$, with the additional relation 
$\hat \Phi_{jwn}=\hat \Phi_{-j -w n}\ e^{i {\pi}n}$.
In this case the $C_\ell$s become
\[
C_\ell \propto
\sum_{jwn} 
\frac{{\cal P}(k)}{k^3} 
\frac{j_\ell(\Delta \eta k)^2}
{2 \ell + 1} 
\sum_{m = -\ell}^{\ell}
|Y_{\ell,m}({\bf \hat k})|^2
( 1 + e^{i \pi (n+m)}).
\]

Another possible compact space identifies opposite faces 
of the parallelepiped with 
one face rotated by $\pi/2$.
The discrete eigenmodes are
$\vec k=2\pi(j / h, w/b,n/4c)$, with the additional relations
$\hat \Phi_{jwn}=\hat \Phi_{w -j n}e^{in\pi/2}
=\hat \Phi_{-w -jn}e^{in\pi}=\hat \Phi_{-w j n}e^{i 3n\pi/2}$.
The $C_\ell$s are given by
\ba
C_\ell & \propto
\sum_{jwn} 
\frac{{\cal P}(k)}{k^3} 
\frac{j_\ell(\Delta \eta k)^2}
{2 \ell + 1} \sum_{m = -\ell}^{\ell}
|Y_{\ell,m}({\bf \hat k})|^2 \\ \nonumber
&( 1 + e^{i (n+m){\pi}/2} +
e^{i (n+m){\pi}} + e^{i 3(n+m){\pi}/2}).
\ea

The last parallelepiped 
is unique among the flat topologies in that it
has a fundamental domain of volume $2 h b c$ and is thus
a ``double'' parallelepiped.
It is described by the following identifications \cite{wolf}:
Translate along $x$ and then 
rotate around $x$ by $\pi$ 
so that $(x,y,z)\rightarrow (x+h,-y,-z)$. 
Next, translate along $y$ and $z$, then rotate around
$y$ by $\pi$
so that $(x,y,z)\rightarrow (-x,y+b,-(z+c))$. 
Finally, translate along $x,y$ and $z$, then rotate around
$z$ by $\pi$
so that $(x,y,z)\rightarrow (-(x+h),-(y+b),z+c)$. 
The discrete spectrum for this space is 
$\vec k=\pi(j / h, w/b,n/4c)$.
With relations as given in LSS
we find
\ba
C_\ell & \propto
\sum_{jwn} 
\frac{{\cal P}(k)}{k^3} 
\frac{j_\ell(\Delta \eta k)^2}
{2 \ell + 1} \sum_{m = -\ell}^{\ell}
Y_{\ell,m}^*({\bf \hat k}) \\ \nonumber
&\big(
Y_{\ell,m}({\bf \hat k}) (1 + e^{i (m+j+w+n)\pi}) \\ \nonumber 
&+ Y_{\ell,-m}({\bf \hat k}) e^{i \ell \pi}
(e^{i (m+j)\pi} + e^{i (w+n)\pi}) \big).
\ea

\begin{figure}
\centerline{\psfig{file=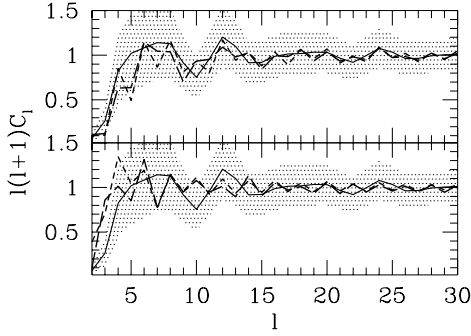,width=2.5in}}
\caption{Normalized $C_\ell$s for the parallelepiped cases.
In the upper panel, the 
$\pi$-twisted torus is short dashed, the $\pi/2$-twisted torus is long dashed,
both with the topology scale equal to the horizon size.
In the lower panel, the multiply-twisted case with $h=b=c=\Delta\eta$ is
dotted, and the multiply twisted case with $V=\Delta\eta^3$ is dashed.
In both plots the solid line is the unit torus with
cosmic variance, provided for reference.}
\label{fig:boxes}
\end{figure}

In Fig.\ \ref{fig:boxes} we plot the normalized $C_\ell$s for each 
of these spaces, taking the topology scale equal to the horizon size.
In the upper panel we see that both the $\pi$-twisted and $\pi/2$
twisted tori have $C_\ell$s that are almost identical to that of the
torus.  
It may seem that the maximum wavelength would be bigger
for twisted spaces since a wave can wrap more than once across the fundamental
domain before completing a full circuit.
However, a closer inspection shows that the relations between eigenmodes 
places a cutoff in these spectra at the same place as in the torus
(LSS).
While it is true that in some sense, there is power at
smaller eigenvalues in the twisted direction, the angular average
discards this asymmetric information.
Here we see that not only is the cutoff
the same for square parallelepipeds, 
but the $C_\ell$s are damped in almost the same
manner.  The harmonics appear to be the distinguishing 
feature between topologies, but unfortunately
fall well within cosmic variances.

The multiply twisted case is somewhat less damped than the other
cases, due to the ``double'' nature of the fundamental domain.
Halving the volume lessens these differences,
although the relations in this topology still allow slightly
larger modes than
the other cases.  This is illustrated in the lower panel of 
Fig.\ \ref{fig:boxes}.

The last two possible compact flat spaces are based on a
hexagonal tiling. In the first of these cases, 
the opposite sides of the hexagon are identified while
in the $z$ direction the faces are
rotated relative to each other by 
$2\pi/3$.  The potential can be written as
	\ba
	\Phi &=&\sum_{n_2 n_3n_z} \hat \Phi_{n_2 n_3 n_z}
	e^{ik_z z}	 
	\times \label{hex}\\
	&\exp &{\left [
	i{2\pi \over h}\left [
	n_2\left ( x -{1\over \sqrt{3}} y \right )
	+n_3\left ( x +{1\over \sqrt{3}} y \right )\right ] \right ]}
	\nonumber
	\ea
with the eigenmodes
$\vec k=2\pi((n_2+n_3) / h, (-n_2+n_3)/b,n_z/3c)$.
The relations on this space (LSS) result in $C_\ell$s given by
\ba\
C_\ell & \propto
\sum_{n_2n_3n_z} 
\frac{{\cal P}(k)}{k^3} 
\frac{j_\ell(\Delta \eta k)^2}
{2 \ell + 1} \sum_{m = -\ell}^{\ell}
|Y_{\ell,m}({\bf \hat k})|^2 \\ \nonumber
& ( 1 + e^{i 2(n_z+m){\pi}/3} + e^{i 4(n_z+m){\pi}/3}).
\ea

The last possibility 
identifies the $z$-faces after rotation by $\pi/3$.
The potential can still be written as (\ref{hex}).
With 
$\vec k=2\pi((n_2+n_3) / h, (-n_2+n_3)/b,n_z/6c)$
and a set of relations among the $\hat \Phi_{\vec k}$,
the $C_\ell$s are given by
\ba
C_\ell & \propto
\sum_{n_2n_3n_z} 
\frac{{\cal P}(k)}{k^3} 
\frac{j_\ell(\Delta \eta k)^2}
{2 \ell + 1} \sum_{m = -\ell}^{\ell}
|Y_{\ell,m}({\bf \hat k})|^2 \\ \nonumber
&( 1 + e^{i (n_z+m){\pi}/3} + e^{i 2(n_z+m){\pi}/3} \\ \nonumber
&+ e^{i (n_z+m){\pi}} + e^{i 4 (n_z+m){\pi}/3} 
+ e^{i 5 (n_z+m){\pi}/3}).
\ea
The volume of both of these topologies is $h^2 c \frac{\sqrt{3}}{2}$.

\begin{figure}
\centerline{\psfig{file=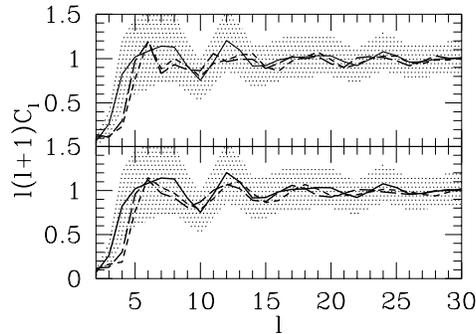,width=2.5in}}
\caption{Normalized $C_\ell$s for the hexagonal cases. The short dashed lines
correspond to $h=c=\Delta\eta$, the long dashed lines to $V=\Delta\eta^3$,
and the solid to the unit torus.  The $2 \pi/3$-twisted 
space appears in the upper panel and the $\pi/3$-twisted space in the 
lower panel.}
\label{fig:hex}
\end{figure}

In Fig.\ \ref{fig:hex} we plot normalized $C_\ell$s for both
hexagonal
spaces with unit volumes and with $h=c=\Delta\eta.$
Again, the damping is quite similar to the torus, despite
the fact that modes in the $\pi/3$ torus must wrap around a full
six times before being associated with the same point. The $C_\ell$s
of unit volume are a somewhat better match.
% as can be understood
%from the fact that the shortest distance across the hexagonal face in 
%this model is equal to $\Delta \eta.$

\section{Angular Correlation Function}

We now turn our attention to the angular correlation function itself.
While $C(\theta)$ can in principle be obtained numerically from the
$C_\ell$s through a Legendre transform,
%\be
%C(\theta) =
%\frac{1}{4 \pi} \sum_{\ell} (2 \ell +1) C_\ell P_\ell(\cos \theta),
%\ee
an analytical expression 
can be obtained 
for the special case of the torus
by carrying out an angular average over the
sky with the angle between $\hn$ and $\hnp$ fixed.
This gives us
\be
C(\theta) \propto
\sum_{jwn} \frac{{\cal P}(k)}{k^3} 
\frac{\sin(2\Delta \eta k \sin (\theta/2)}
          {2\Delta \eta k \sin (\theta/2)}.
\ee

\begin{figure}
\centerline{\psfig{file=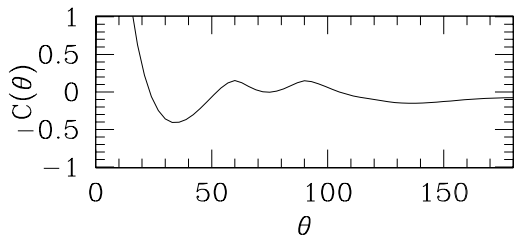,width=2.8in}}
\centerline{\psfig{file=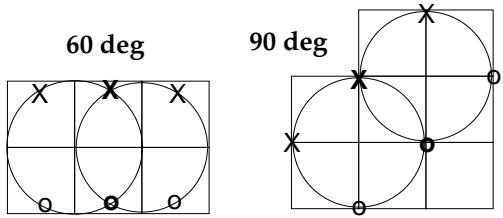,width=2.6in}}
\vskip .1in
\caption{upper: $C(\theta)$ for a torus with the topology scale set equal
to the horizon scale.  lower: Intersections of the last scattering surface
with itself, corresponding to secondary peaks in $C(\theta)$.}
\label{fig:tor1.0}
\end{figure}

In the upper panel of Fig.\ \ref{fig:tor1.0} we plot $C(\theta)$ for
a flat power spectrum.   Note the secondary peaks
at angular separations of 60 and 90 degrees.  These occur because the 
relationship between the topology and the horizon size is such that 
for certain orientations, multiple copies of the same point appear at 
these angular separations.  These double occurrences are a general feature
of all detectable topologies, although the angular separations will vary.  
For the case of $\Delta \eta = h = b = c$, these pairings are illustrated in
the lower panel of Fig.\ \ref{fig:tor1.0}.

While each pairing is completely correlated, the 
averaging of $C(\hn,\hnp)$ over all orientations causes
the secondary peaks in $C(\theta)$ to be dwarfed by the small angle 
correlations.  This suggests that the most accurate determinations of
topology will employ an approach that retains orientation information.
Such methods are a topic of current research 
\cite{costa,css1,css2,bps,ped,lssb}.

\section{Conclusions}

As there are an infinite number of compact open spaces, 
all of which support chaotic flows, many theoretical 
advances must be made before these cases can be approached. 
While we have only considered flat topologies, at least 
some of the 
features we observe, such as discrete harmonics
of the fundamental domain, enhancement due to ringing, and 
multiple peaks in the correlation function, 
can be expected to play a role in any finite universe.
While the absence of large-scale damping places a 
weak lower bound on the minimum topology length,
the angular power spectrum is in general 
a poor measure of topology.
Not only does it discard vital anisotropic information,
but cosmic variance prevents it from discriminating a
hexagonal prism from a hypertorus.

More inventive methods have been suggested such as
a search for circles in the sky 
\cite{css1,css2}, 
pattern formation 
\cite{lbbs,lssb}, 
or a method of images \cite{bps}.
However we look for these features,
topology is as important 
 as the curvature or the primordial power spectrum
in determining the structure of the
microwave background.
Just as curvature may be observable with a new generation of
experiments, so too may be 
the global structure of the universe.

\section*{Acknowledgments}

This research has been supported in part by grants from NASA
and DOE.
We appreciate  valuable input from 
J.R. Bond, N. Cornish, P. Ferreira, K. Gorski, D. Pogosyan,
T. Souradeep, D. Spergel and G. Starkman.  
E.S. is supported by the NSF.


\begin{thebibliography}{}

\bibitem[Bond, Pogosyan, \& Souradeep 1997]{bps}  
Bond, J.R., Pogosyan, D., \& Souradeep, T. 1997,
Proceeding of the 18th Texas Symposium on Relativistic
Astrophysics, ed. Olinto, A., Reieman, J., \& Schrammn D.
(World Scientific, Singapore)

\bibitem[Cornish, Spergel, \& Starkman 1996]
{css1} N.J. Cornish, D. Spergel and G. Starkman 1996, 
Phys. Rev. Lett., 77, 215

\bibitem[Cornish, Spergel, \& Starkman 1997]{css2} 
Cornish, N.J., Spergel D., \& Starkman G. 1997, preprint 
(astro-ph/9708225)

\bibitem[de Oliveria-Costa, Smoot, \& Starobinsky  1996]{costa} 
de Oliveria-Costa, A., Smoot, G.F., \&  Starobinsky, A.A. 1996 
ApJ, 468, 457

\bibitem[Ferreira \& Magueijo 1997]{ped} 
Ferreira, P. \& Magueijo, J. 1997, Phys. Rev D, 56,
4578

\bibitem[Garnavich et al. 1998]{open1} 
P. Garnavich et al., 1998 ApJ, 493, L53 

\bibitem[Gott 1980]{flat1}
Gott, J.R. 1980, MNRAS, 193, 153

\bibitem[Lachieze-Rey \& Luminet 1995]{lum1} 
Lachieze-Rey, M. \& Luminet, J.P. 1995, Phys. Rep., 254,
135

\bibitem[Lehoucq, Lachieze-Rey, \& Luminet 1996]{lum2}
Lehoucq, R., Lachieze-Rey, M., \& Luminet, J.P. 1996, preprint 
(gr-qc/9604050)

\bibitem[Levin, Barrow, Bunn, \& Silk 1997]{lbbs}  
Levin, J. Barrow, J.D. Bunn, E.F., \& Silk, J. 1997,
Phys. Rev. Lett., 79, 974 

\bibitem[Levin, Scannapieco, \& Silk 1998]{us}  
Levin, J., Scannapieco, E. , \& Silk, J.  1998a, Phys. Rev. D58 103516.

\bibitem[Levin et al. 1998]
{lssb}  Levin, J., Scannapieco, E., de Gasperis, G., Silk J., \& Barrow, J.D.
1998b, to appear in Phys Rev. D58.

\bibitem[Spinrad et al.\ 1997]{open2} 
Spinrad, H., Dey, A., Stern, D., Dunlop, J., Peacock, J., Jimenez, R., 
Windhorst, R. 1997, ApJ, 484, 581

\bibitem[Starobinsky 1993]{star}
Starobinsky, A. A. 1993, JETP Lett. 57 622

\bibitem[Stevens, Scott, \& Silk 1993]{sss}  
Stevens, D., Scott, D., \& Silk, J. 1993,
 Phys. Rev. Lett., 71, 20

\bibitem[Sokolov 1993]{flat2}
Sokolov, I.Y. 1993, JETP Lett., 57, 617

\bibitem[Wolf 1967]{wolf} Wolf, J.A. 1967, Spaces of Constant Curvature
(Wilmington: Publish or Perish, Inc.)


\end{thebibliography}
\end{document}